\documentclass[twocolumn, prl, amssymb, superscriptaddress, aps, 
preprintnumbers,
amsmath,
floatfix
]{revtex4-1} 
\usepackage{multirow}
\usepackage{isotope}
\usepackage{graphicx}
\usepackage{color}
\usepackage{braket}
\usepackage{hyperref}

\begin{document}

\title{Deep-learning approach for the atomic configuration interaction problem\\on large basis sets}

\author{Pavlo Bilous}
\email{pavlo.bilous@mpl.mpg.de}
\affiliation{Max Planck Institute for the Science of Light, Staudtstra{\ss}e 2, 91058 Erlangen, Germany}
\affiliation{Max Planck Institute for Nuclear Physics, Saupfercheckweg 1, 69117  Heidelberg, Germany}
\author{Adriana P\'alffy}
\affiliation{Institute of Theoretical Physics and Astrophysics, University of W\"urzburg, Am Hubland, 97074 W\"urzburg,  Germany}
\affiliation{Max Planck Institute for Nuclear Physics, Saupfercheckweg 1, 69117  Heidelberg, Germany}
\author{Florian Marquardt}
\affiliation{Max Planck Institute for the Science of Light, Staudtstra{\ss}e 2, 91058 Erlangen, Germany}
\affiliation{Department of Physics, Friedrich-Alexander-Universit\"at Erlangen-N\"urnberg, 91058 Erlangen, Germany}

\date{\today}

\begin{abstract}

High-precision atomic structure calculations require accurate modelling of electronic correlations typically addressed via the configuration interaction (CI) problem on a multiconfiguration wave function expansion.  The latter can easily become challenging or infeasibly large even for advanced supercomputers.  Here we develop a deep-learning approach which allows to preselect the most relevant configurations out of large CI basis sets until the targeted energy precision is achieved.  The large CI computation is thereby replaced by a series of smaller ones performed on an iteratively expanding basis subset managed by a neural network. While dense architectures as used in quantum chemistry fail, we show that a convolutional neural network naturally accounts for the physical structure of the basis set and allows for robust and accurate CI calculations.  The method was benchmarked on basis sets of moderate size allowing for the direct CI calculation,  and further demonstrated on prohibitively large sets where the direct computation is not possible.

\end{abstract}

\maketitle

The precise knowledge of atomic structure is indispensable for frequency standards in metrology,  spectral analysis in astrophysics, understanding of nuclear phenomena involving atomic electrons, or investigations of physics beyond the standard model,  e.g space and time variation of fundamental constants~\cite{Fischer_Review_JPhysBAtMolOptPhys_2016}. 
\textit{Ab initio} atomic structure calculations are the scope of high performance  codes that provide a wide range of electronic properties of atoms and ions,  such as energy levels,  radiative transition rates,  $g$-factors or hyperfine structure constants.  The practical difficulty arises from  many-body effects when considering atoms or ions with high atomic number $Z$ and many electrons. The electronic correlations are typically tackled by the configuration interaction (CI) method based on the multiconfiguration wave function expansion $\ket{\Psi} = \sum_\alpha c_\alpha \ket{\Phi_\alpha}$ with unknown coefficients $c_\alpha$ obtained as a solution of the Hamiltonian diagonalization problem $\hat{H}\ket{\Psi} = E \ket{\Psi}$~\cite{Grant_book_2007}.  The size of the involved basis set $\{\ket{\Phi_\alpha}\}$ can easily become challenging even for state-of-the-art parallelized codes running on supercomputer systems, see e.g.  recent calculations of electronic energy levels in $\isotope{Th}^{35+}$~\cite{Bilous_HCI_PRL_2020,  Porsev_Th35+_2021, Peik_BigReview_Th_2021},  $\isotope{Ir}^{17+}$~\cite{Cheung_ScalCodesSym_2021,  Kozlov_HCI_RevModPhys_045005_2018} or $\isotope{Fe}^{16+}$~\cite{Cheung_ScalCodesSym_2021,  Kuehn_FeAstro_PRL_225001_2020}.
 
Instead of computations on the full basis,  ``selected CI'' methods were applied to atomic and molecular systems using selection criteria based on perturbation theory~\cite{Harrison_PTCI, Dzuba_CI+MBPT_PRA_1996} or the Monte-Carlo approach~\cite{Greer_MCCI, GREER1998181}.  However,  these methods become inefficient for large basis sets,  since perturbation theory still requires computations on the entire basis,  whereas the random selection completely disregards the properties of the basis states.  Fortunately,  application of machine learning techniques has lead in recent years to significant progress in selected CI in quantum chemistry~\cite{Coe_MLCI_JChemTC_2018,  Jeong_ALCI_JChemTC_2021, Chembot, RLCI}.  For importing this success in the field of large-scale atomic computations,  the neural network (NN) approach would be the first choice due to its established scalability and flexibility~\cite{Goodfellow-et-al-2016}.  Used in the active learning algorithm as presented in Refs.~\cite{Coe_MLCI_JChemTC_2018,  Jeong_ALCI_JChemTC_2021},  it would allow us to iteratively construct a compact wave function that delivers accurate observables without the computational effort on the full basis.  However, we show here that the usual dense architectures as applied in Refs. ~\cite{Coe_MLCI_JChemTC_2018,  Jeong_ALCI_JChemTC_2021} lack sufficient accuracy and often fail.  In this work we apply instead a convolutional NN (CNN) --- the architecture well known from image recognition applications~\cite{Goodfellow-et-al-2016,  NIPS2012_c399862d}.  We demonstrate that this is the natural choice taking into account the physical structure of the basis states of atomic systems which leads to robustness of the approach and a strong improvement of the computational results.

In this Letter,  we develop an efficient deep-learning approach to iteratively construct a compact approximative wave function for high-$Z$ atoms and ions with many electrons.  We address the problem in the coupled basis of configuration state functions (CSF)~\cite{Fischer_Review_JPhysBAtMolOptPhys_2016,  Grant_book_2007} characterized by electronic orbital occupations,  and the angular momenta couplings within and between the orbitals.  Typically,  the physical properties of a CSF are determined predominantly by a few orbitals (different for each CSF).  The others form a ``background'' consisting of low-energy fully occupied and high-energy empty orbitals.  In analogy to image recognition applications,  the applied CNN efficiently suppresses this background and highlights the ``useful image'' of the physically relevant orbitals in each CSF.  We demonstrate that this natural choice accounting for the physical structure of CSFs leads to significant improvements in comparison to NNs of the usual dense type.  Our solver of the CI problem is based on an iterative scheme employing the CNN together with the General Relativistic Atomic Structure Package GRASP2018~\cite{grasp2018}.

\begin{figure}[ht!]
\centering
\includegraphics[width=0.5\textwidth]{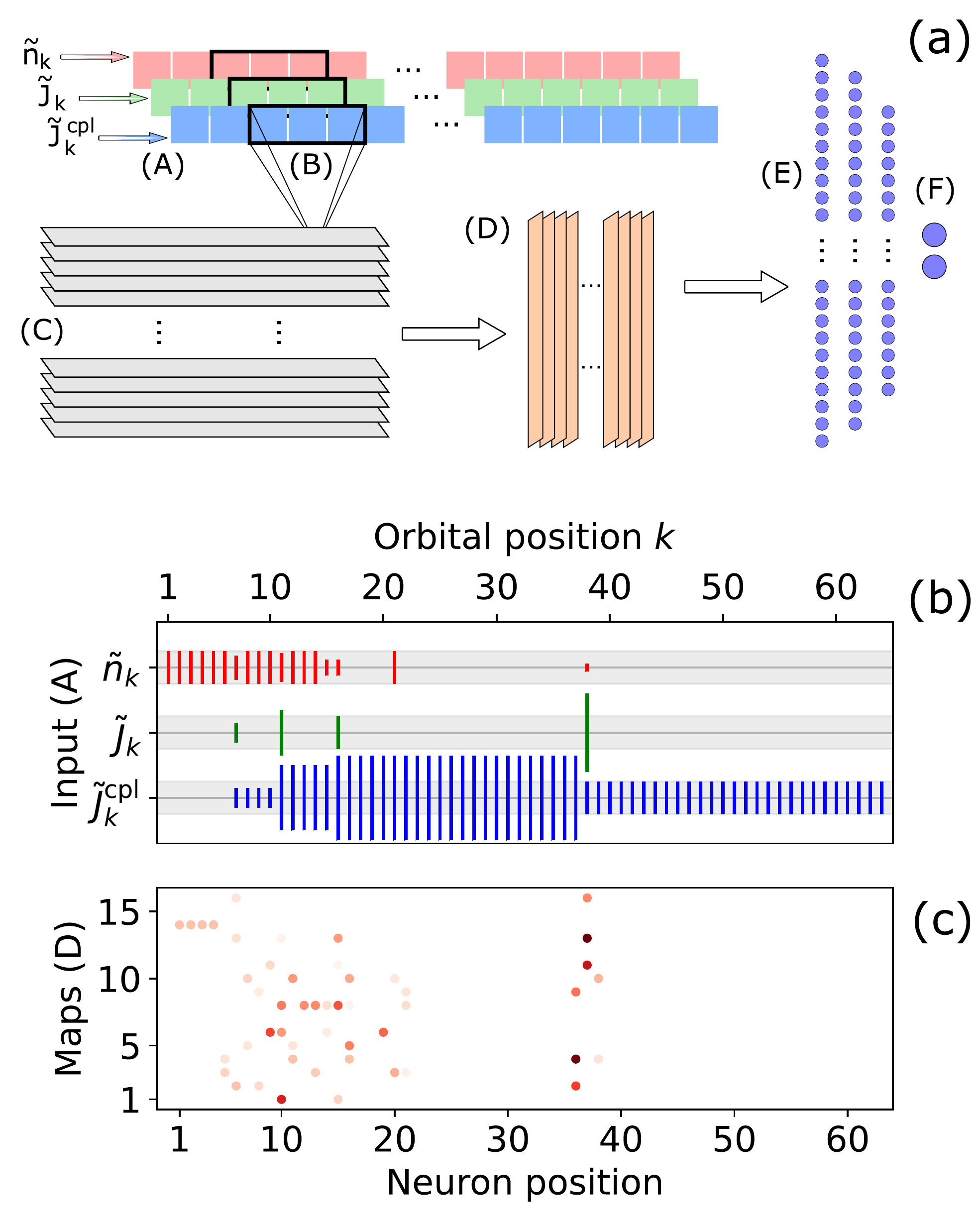}
\caption{(Color online) (a): NN architecture used in the present work; (b): Color representation of the exemplary CSF from the set $\mathrm{SD}^*(3p, 9h)$ of the Re atom as the NN input~(A); (c): The background suppression observed in the neurons of the feature maps~(D) of the NN.  See the text and Supplemental Material \cite{SupplMat} for details.}
\label{cnn_imgrec_fig}
\end{figure}

In the wave function expansion $\ket{\Psi} = \sum_\alpha c_\alpha \ket{\Phi_\alpha}$ with $N$ electronic orbitals,  the CSFs $\ket{\Phi_\alpha}$ are uniquely characterized by the set of $3N$ quantum numbers generically denoted as $\alpha$. For each $k$-th orbital,  they consist of its population $n_k$,  the total angular momentum of its electrons $J_k$,  and the angular momentum $J_k^\text{cpl}$ representing the coupling of $J_k$ and $J_{k-1}^\text{cpl}$ \cite{Edmonds,SpringerHandbookAMO} (we assume $J_1^\text{cpl} = 0$).  We normalize the populations of the orbitals $n_k$ to their maximal capacity,  and the angular momenta $J_k$ and $J_k^\text{cpl}$ to the total angular momentum of the considered energy level (see also Supplementary Material \cite{SupplMat} for the CSF basis construction).  The three classes of the input data $\tilde{n}_k$,  $\tilde{J}_k$,  $\tilde{J}_k^\text{cpl}$ obtained in this way are interpreted as color channels of a 1D convolutional input layer.  In Fig.~\ref{cnn_imgrec_fig}b we show graphically this color representation for an exemplary CSF (belonging to the set $\mathrm{SD}^*(3p, 9h)$ of the Re atom ground state,  see below).  The value of each parameter is encoded by the length of the corresponding vertical bar,  whereas the grey horizontal strips indicate the unity bar length.  We consider here the natural ordering of the orbitals which is default in GRASP2018.

A network architecture that we found to work efficiently and which we focus on here is shown in Fig.~\ref{cnn_imgrec_fig}a.  The input layer  (A)  consists of 3 color channels (see Fig.~\ref{cnn_imgrec_fig}b) of size $N$.  The input is processed with a filter kernel of size 3 (B) resulting in $96$ feature maps (C) each of size $N-2$.  The latter are mapped to 16 feature maps (D) of size $N-2$ by application of a filter kernel of size 1 (thus representing a purely local transformation).  The CNN part (D) is monitored for observation of the background suppression effect which we show in Fig.~\ref{cnn_imgrec_fig}c and discuss in detail further on.  The obtained output of $16 \times (N-2)$ values is then flattened and forwarded to a network of 4 dense layers (E) with 150, 120,  90 and 2 neurons, respectively.  The ReLU activation function was used throughout the NN apart from the two-neuron output layer (F), where the softmax function is applied yielding the probabilities of the CSF to be ``important'' or ``unimportant'' (see below).  The categorical cross-entropy was chosen as the loss function.  The NN is trained on batches using the Adam optimization algorithm with early stopping based on the classification accuracy evaluated on a validation set (20\% of data excluded in advance from the training set).  The described NN functionality was implemented using the Python library Keras~\cite{chollet2015keras, gulli2017deep} with TensorFlow~\cite{abadi2016tensorflow} in the backend. 

The described NN is employed in an iterative active-training algorithm based on the scheme from Refs. ~\cite{Coe_MLCI_JChemTC_2018, Jeong_ALCI_JChemTC_2021}.  Each CSF is either important or unimportant: its weight in the CI wave function $w_\alpha = |c_\alpha|^2$ exceeds or not some cutoff value $w^0$ chosen in advance.  CSFs are included in the CI expansion iteratively in relatively small portions based on the NN-prediction of their importance.  The diagonalization using GRASP2018  yields the energy and the coefficients $c_\alpha$.  The latter are used for a feedback and additional training of the NN,  whereas the energy is monitored in order to stop the computation when the targeted precision is achieved.  In each iteration, CSFs which turned out to be unimportant are excluded from the CI expansion,  but are considered again in later iterations.  Instead of using a fixed cutoff as in Refs. ~\cite{Coe_MLCI_JChemTC_2018, Jeong_ALCI_JChemTC_2021},  we use a running cutoff taking at the $i$-th iteration a new value $w^i < w^{i-1} < \ldots < w^1 < w^0$.  This approach is crucial for avoiding energy convergence to an unwanted value which does not correspond to the full set, but to a smaller set of CSFs having weights that exceed the fixed cutoff value.  This point was also observed in Ref.~\cite{Chembot}. 

The NN needs feedback not only on the selected but also on rejected CSFs.  Therefore,  we include in the CI expansion the same amount of randomly picked disregarded CSFs as the selected ones in every iteration (but the last one since no NN training follows).  Most of these balancing CSFs are automatically excluded in the next iteration due to their small weights.  Before feeding into the NN, the training data are reshuffled for avoiding the ordering bias due to the CSF construction procedure.  Some important CSFs for a particular electron configuration are known from the start and should always be included in the CI set.  They form the primary subset and we do not expose them to the NN at any stage.  We have checked that inclusion of the primary subset in the training set does not bring any improvement for the method.  At the starting point,  the NN is trained on a random selection of CSFs from the considered basis set (excluding the primary subset).  At the same time, the distribution of these CSFs over their weights is used to choose the running cutoff values $w^i$. Throughout this work we  use 1\% of CSFs for this starting iteration. 

For demonstration of our approach, we choose the case of neutral $\isotope[187]{Re}$ and $\isotope[187]{Os}$ atoms modelled by the Dirac-Hartree-Fock Hamiltonian~\cite{Grant_book_2007} and calculate energies of their ground states with the electronic configurations $[\isotope{Xe}]\,4f^{14}\,5d^5\,6s^2$ and $[\isotope{Xe}] \,4f^{14}\,5d^6\,6s^2$,  respectively.  These energies have been recently evaluated with GRASP2018 to extract the  $\beta$-decay energy of the $\isotope[187]{Re}$ nucleus from experimentally determined masses of $\isotope[187]{Re}^{29+}$ and $\isotope[187]{Os}^{29+}$~\cite{Filianin_PRL_072502_2021}.  The basis sets in Ref. \cite{Filianin_PRL_072502_2021} contain states stemming apart from the main configuration of the considered level also from additional configurations obtained by allowing for electronic excitations from the main configuration.  Single (S) and double (D) excitations from the filled orbitals down to $3p$ to the vacant (virtual) orbitals up to $9h$ were considered in~\cite{Filianin_PRL_072502_2021} resulting in over 90 million CSFs (see Supplemental Material \cite{SupplMat} for more information on the basis construction). We denote these sets here as $\mathrm{SD}(3p, 9h)$.  Due to the prohibitively large basis set size,  the authors of Ref.  \cite{Filianin_PRL_072502_2021} had to preselect about 5 million most important CSFs by evaluating transition and ionization energies and fitting them to experimental values~\cite{Chunghai-private}. 

For benchmarking we apply our method to a smaller $\isotope[187]{Re}$ basis set $\mathrm{SD}^*(3p, 9h)$ of 4,267,362 CSFs in which only part of the double excitations are allowed, with the restriction that each virtual non-relativistic orbital can be either doubly occupied or empty.  The moderate size of $\mathrm{SD}^*(3p, 9h)$  allows for comparison of our approach to direct GRASP2018 computations. The primary subset consists of 37,220 CSFs constructed from SD excitations to the valence orbitals and S excitations to the virtual orbitals.  The radial electronic wave functions are obtained with GRASP2018 in advance on the primary CSF set using the layer-by-layer procedure as described in Ref.  \cite{GRASP2018_guide}.  Table~\ref{NNres_table} shows the results obtained in each iteration: the energy $E^\mathrm{part}$ on the current partial CI set with respect to the exact value $E^\mathrm{all} = -454,661.1637\text{ eV}$ (obtained separately in a direct calculation) and the number of CSFs in the GRASP2018 run.  We note that for CI on a partial basis the energies always satisfy $E^\text{part} > E^\mathrm{all}$~\cite{Szabo_ModnQChem_2012}.  The iterations are labelled by $\log_{10} w^i$ where $w^i$ is the running cutoff value at the $i$-th iteration.  The starting iteration on 1\%  randomly chosen CSFs is also represented in Table~\ref{NNres_table} in the row labelled as ``Init.''.  After the very last iteration,  CSFs unimportant with respect to the value $\log_{10} w^i=-11.6$ as calculated by GRASP2018 are excluded from the CI wave function yielding the final CI set with 729,451 instead of 755,766 CSFs.  The latter step is helpful if the obtained wave function is intended for further calculations on the state,  e.g.  refining of the radial wave functions or evaluation of QED corrections and isotope shifts.  Both the direct and the NN-supported computations could be carried out in a few hours on a few hundred cores.  The peak memory and disk space consumption which are the bottleneck in larger GRASP2018 computations (see Supplemental Material \cite{SupplMat}) could be decreased in this case from a TB to a few hundred GB.

\begin{table}[ht!]
\begin{tabular}{|c|c|c|c|}
\hline
$\log_{10} w^i$  & CSFs in GRASP & $E^\text{part} - E ^\text{all}\,  (\text{meV})$ \\
\hline\hline
Init. 		&  79,521  & 17,223.3 	   \\ \hline
-8.6		&	 178,901  & 6,431.2		 \\  \hline
-9.2 		& 364,562  & 802.9		 \\  \hline
-9.8 		& 515,289  & 140.3		 \\  \hline
-10.4 	& 723,540   &  31.4 		 \\  \hline
-11.0 	& 755,766  & 6.4 	 \\
\hline
\end{tabular}
 \caption{Results of approximate energy calculations on the $\mathrm{SD^*}(3p, 9h)$ basis set for the $\isotope{Re}$ atom ground state using our deep-learning-based approach.  Iterations are labelled by $\log_{10} w^i$ where $w^i$ is the running cutoff value at the $i$-th iteration.  The row labelled as ``Init.'' represents the initial iteration on 1\% randomly chosen CSFs.}
 \label{NNres_table}
\end{table}

\begin{figure}[ht!]
\centering
\includegraphics[width=0.45\textwidth]{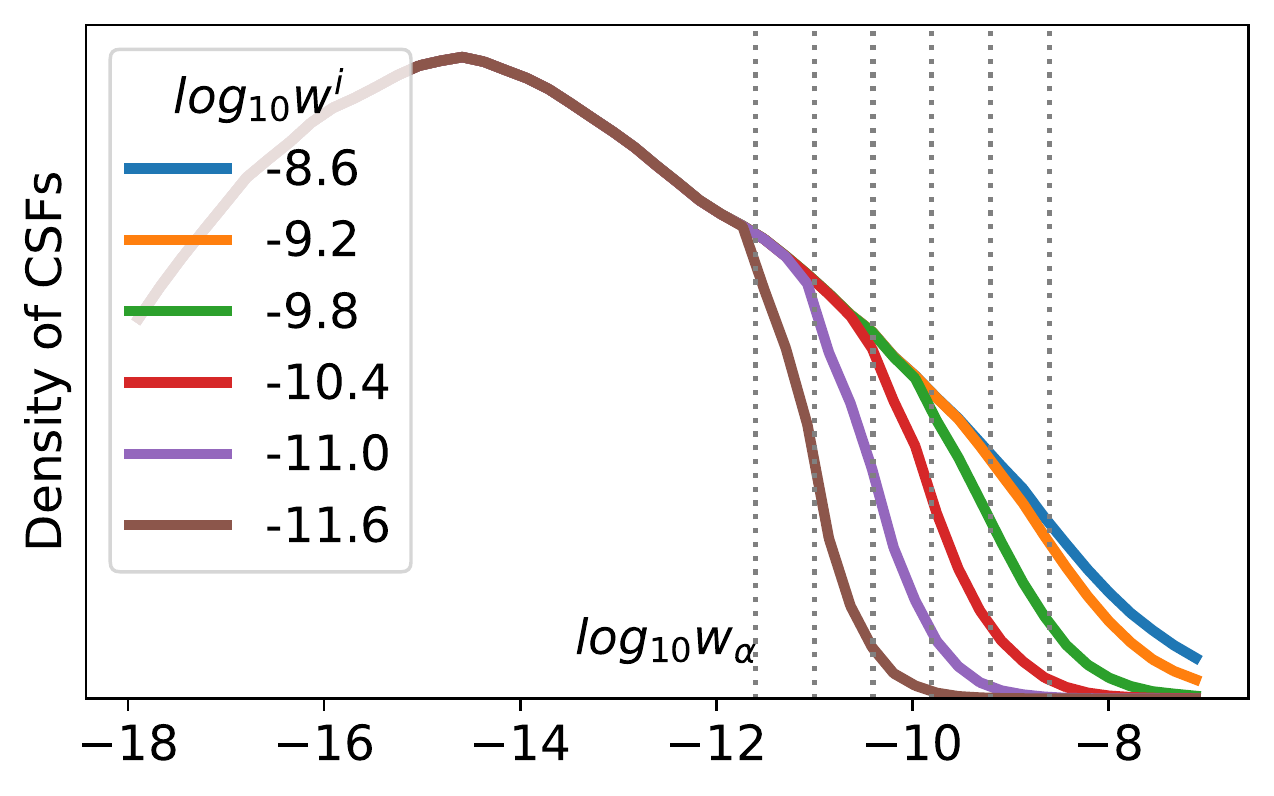}
 \caption{(Color online) The distribution of CSF from the set $\mathrm{SD^*}(3p, 9h)$ not (yet) included in the CI expansion of the wave function as a function of their respective weights $\log_{10} w_\alpha$ for each iteration. The distributions are normalized with respect to the total size of the $\mathrm{SD^*}(3p, 9h)$ set.
 The iterations are labelled by the cutoff  values $\log_{10} w^i$ which are additionally illustrated by the vertical dotted lines. 
  }
\label{NNres_fig}
\end{figure}

Fig.  \ref{NNres_fig} illustrates graphically the growth of the CI expansion of the wave function for the considered example.  We plot the number of CSFs from the set $\mathrm{SD^*}(3p, 9h)$ not (yet) included in the CI expansion as a function of the weight  $\log_{10} w_\alpha$ for each iteration immediately after the unimportant CSFs are excluded.  The distribution is normalized with respect to the total size of the $\mathrm{SD^*}(3p, 9h)$ set and  the weights $w_\alpha$ are taken from the full GRASP2018 calculation.  In each iteration, CSFs  are included in the CI expansion (and thus removed from the depicted distributions) from the right.  The right edge of the distributions is not completely sharp,  meaning that not all CSFs important with respect to the current cutoff are included in the WF set. The NN selection ensures that  the slope becomes stable in the first iterations and moves then from right to left. 
 
\begin{figure}[ht!]
\centering
\includegraphics[width=0.5\textwidth]{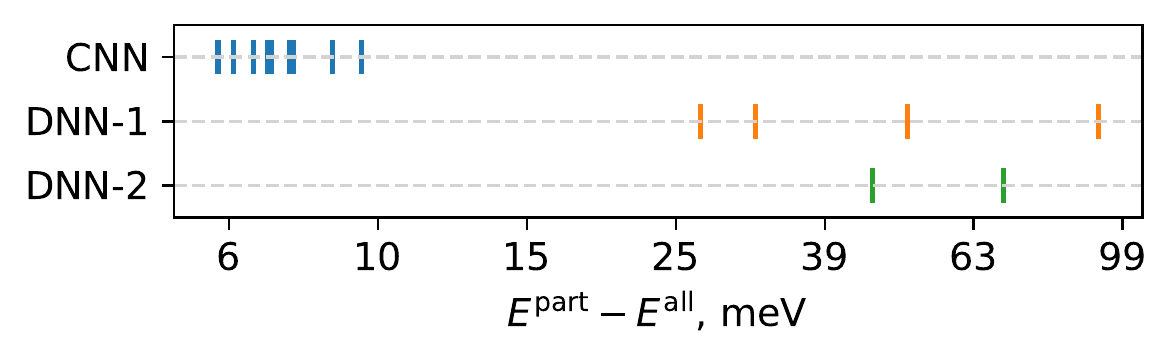}
\caption{Deviation of the final energy on the partial set $E ^\text{part}$ with respect to the ``full'' energy $E ^\text{all}$ obtained using the CNN (10 runs) and dense networks DNN-1 and DNN-2 (5 runs each).   Note logarithmic scale on the horizontal axis.  The missing DNN runs have failed.  See text for further explanations.}
\label{cnn_dnn_fig}
\end{figure}

We have performed the same computation replacing the CNN by a usual dense NN.  We considered two dense NN architectures: DNN-1 is the dense part (E)---(F) of the applied CNN (see Fig.~\ref{cnn_imgrec_fig}a); DNN-2 has 3 hidden layers with 192,  384,  192 neurons, respectively,  possessing in total a similar number of trainiable parameters as the considered CNN.  In Fig.~\ref{cnn_dnn_fig} we show the final energy $E^\text{part} - E ^\text{all}$ in meV obtained in a few computation repetitions using the CNN and dense networks DNN-1 and DNN-2.  Often,  the DNNs fail,  and  this takes place already in the first iteration,  which is more challenging than the subsequent iterations from the point of view of the training data structure.  Indeed,  the very first training set is a random selection yielding a small fraction of important CSFs and is thus strongly disbalanced.  In contrast to the DNNs,  our CNN is more robust against this disbalance and failed only a few times in hundreds of runs.  In case the DNNs do cope with the first iteration,  they are still strongly outperformed by the CNN.  The processing of the input data using a kernel in the CNN plays a two-fold role: (a) Application of the same weights along the input neurons (independent of the orbital ordering); and (b) revealing data mutual dependencies for the neighbouring input neurons (sensitive to the orbital ordering).  A careful analysis on how these mechanisms contribute to the performance of the CNN reveals that (a) plays the most important role  (see  Supplemental Material \cite{SupplMat}).

The  CNN tends to treat the fully occupied low-lying orbitals as well as the completely vacant high-energy orbitals as a common background.  This background is suppressed and the remaining ``useful image'' corresponding to the physically relevant partially occupied orbitals is highlighted.  We demonstrate this effect in Fig.~\ref{cnn_imgrec_fig}c which shows the values of the 62 neurons in each of the 16 feature maps (D) for the exemplary CSF.  These neurons are in positional correspondence with the 64 input neurons (A).  The values in (D) are plotted at the moment when the computation is completed and the NN is in its final state.  The color intensity of the red dots indicates the (always non-negative) values normalized to the maximal value in all the feature maps (higher intensity corresponds to a larger value).  Almost all neurons in the region of the common background to  the left and to the right turn out to have zero values after the NN training.  Further discussion of this effect and more examples can be found in the Supplemental Material \cite{SupplMat}.

We switch now to calculations on the large basis sets $\mathrm{SD}(3p, 9h)$ for the $\isotope{Re}$ and $\isotope{Os}$ neutral atoms relevant for the determination of the $\isotope[187]{Re}$ $\beta$-decay energy in Ref.  \cite{Filianin_PRL_072502_2021}.  These calculations involve basis sets of over 90 million CSFs each and cannot be performed directly using GRASP2018.  However,  it is sufficient to retain for each basis set only the most important CSFs that deliver a 1 eV precision for the calculated energy.  Using our deep-learning approach, we could achieve the targeted accuracy in a few days by performing partial GRASP2018 runs on up to about 5 million CSFs which required 5 TB of memory and 7 TB of the disk space.  The primary CSF subsets were constructed as in the previous example resulting in 37,220 and 32,660 CSFs for the $\isotope{Re}$ and $\isotope{Os}$ atom,  respectively.  The radial electronic wave functions were obtained on the primary CSF sets in the same way as before.  In the combined Table~\ref{NNres_big_table} for the $\isotope{Re}$ and  $\isotope{Os}$ atoms we show the energies and the basis set size at the diagonalization stage in each iteration.  From the convergence pattern it is seen that the energy values obtained in the last iteration satisfy our precision target.  We carried out additional verifications by running the computation on other cutoffs and made sure that they lead to the same energy value within the required accuracy.  The binding energies of the $\isotope{Re}$ and  $\isotope{Os}$ neutral atoms are thus $E_\mathrm{Re}^\text{atom} = 454,703.55\text{ eV}$ and $E_\mathrm{Os}^\text{atom} = 470,036.60\text{ eV}$, respectively. 

\begin{table}[ht!]
\begin{tabular}{|c||c|c||c|c|}
\hline
 &  \multicolumn{2}{c||}{Re} & \multicolumn{2}{c|}{Os} \\\hline\hline
\multirow{2}{*}{$\log_{10} w^i$} & \multirow{2}{*}{CSFs} & $E^\text{part},  \text{ eV}$ & \multirow{2}{*}{CSFs}  & $E^\text{part},  \text{ eV}$ \\
 &  & 454,000 + & & 470,000 + \\
\hline\hline
Init. & 971,011 & 644.55    &   985,571 & -23.64 \\\hline
-8.0 & 578,018 & 668.16 &   628,961 & 9.07 \\\hline
-8.5 & 1,609,943 & 684.00  &  972,374 & 25.60 \\\hline
-9.0 & 2,055,985 & 697.82  &   1,345,026 & 33.04  \\\hline
-9.5 & 2,550,922 & 701.65  &  2,046,765 & 35.71  \\\hline
-10.0 & 3,607,689 & 702.97  &  2,397,010 & 36.36  \\\hline
-10.5 & 4,028,106 & 703.55  &   3,185,458 & 36.60  \\\hline
\end{tabular}
 \caption{Results of approximate energy calculations on the $\mathrm{SD}(3p, 9h)$ basis set for the ground state of the $\isotope{Re}$ and $\isotope{Os}$ atoms using our deep-learning-based approach.  Iterations are labelled by $\log_{10} w^i$ where $w^i$ is the running cutoff value at the $i$-th iteration.  The row labelled as ``Init.'' represents the initial iteration on 1\% randomly chosen CSFs.}
 \label{NNres_big_table}
\end{table}

In order to compare our results with Ref. \cite{Filianin_PRL_072502_2021} where the electronic binding energy differences $\delta E = E^\text{atom} -  E^\text{ion}$ between a neutral atom and a 29+ ion for Re and Os were provided,  we evaluate the energies of the $\isotope{Re}^{29+}$ and $\isotope{Os}^{29+}$ ions on the basis sets $\mathrm{SD}(3p, 9h)$.  Since these consist of only 53,885 and 2,455,449 CSFs,  respectively,  we carry out the GRASP diagonalization on the full sets directly.  The radial wave functions for the ions were obtained using the layer-by-layer procedure \cite{GRASP2018_guide} on the full set $\mathrm{SD}(3p, 9h)$ for the $\isotope{Re}^{29+}$ ion and on a partial set constructed as a union $\mathrm{SD}(3p, 5g) \cup \mathrm{SD^*}(3p, 6h) \cup \mathrm{S}(3p, 9h)$ for the  $\isotope{Os}^{29+}$ ion.  The obtained ion energies are $E_\mathrm{Re} ^\text{ion}= -443,804.16\text{ eV}$ and $E_\mathrm{Os} ^\text{ion}= -459,083.43\text{ eV}$.  Based on the calculated atom and ion energies,  we find $\delta E_\mathrm{Re} = -10,899.39\text{ eV}$ and $\delta E_\mathrm{Os} = -10,953.17\text{ eV}$,  which agree with the values $\delta \widetilde{E}_\mathrm{Re} = -10,894.5 \pm 25.9 \text{ eV}$ and $\delta \widetilde{E}_\mathrm{Os} = -10,947.9 \pm 24.6 \text{ eV}$ from Ref. \cite{Filianin_PRL_072502_2021}.  For the difference $\Delta E = \delta E_\mathrm{Re} - \delta E_\mathrm{Os}$ relevant for the computation of the $\isotope[187]{Re}$ $ \beta$-decay energy,  we obtain $\Delta E = 53.78 \text{ eV}$,  whereas in Ref. \cite{Filianin_PRL_072502_2021} the value $\Delta \widetilde{E} = 53.4 \pm 1.0 \text{ eV}$ was reported.  In this way,  our approach allowed to achieve the same results as in Ref. \cite{Filianin_PRL_072502_2021} without relying on additional experimental information which is not always available.  We note that the individual atom and ion energies change upon inclusion of the QED corrections and further improvement of the radial wave functions.  At the same time, the obtained value $\Delta E$ does not change significantly due to the cancellation effects originating from similarity of the $\isotope{Re}$ and $\isotope{Os}$ electronic shells --- a fact observed also in Ref. \cite{Filianin_PRL_072502_2021}.

In conclusion,  we have developed a deep-learning-based approach for atomic CI calculations with large CSF basis sets.  The method is based on the iterative scheme described in Refs.~\cite{Coe_MLCI_JChemTC_2018,  Jeong_ALCI_JChemTC_2021} allowing to replace a large CI computation by a number of relatively small ones.  The approximative basis set is expanded iteratively until the targeted energy precision is achieved.  The usual dense architectures as used in Refs.  \cite{Coe_MLCI_JChemTC_2018,  Jeong_ALCI_JChemTC_2021} are not compatible with the CSF physical structure and proved to be unreliable by either yielding low precision results or failing already in the first iteration.  In this work we applied an NN of the convolutional type combined with the ``color'' representation of the CSFs --- a natural choice leading to robustness of the method and a strong improvement in precision.  The method was benchmarked on CSF basis sets of moderate size allowing for the direct CI calculation,  and demonstrated on prohibitively large sets where the direct computation is infeasible.  Analogously to image recognition where CNNs are commonly applied~\cite{Goodfellow-et-al-2016,  NIPS2012_c399862d},  our NN recognizes and suppresses the background corresponding to the fully occupied low-lying and the vacant high-energy orbitals, whereas the ``useful image'' related to the physically relevant partially filled orbitals is highlighted.  
We believe that this approach can be useful also in other areas where the CI method is applied.  The code for NN-supported GRASP computations is available in Ref.  \cite{neural_grasp_code} and can be used with the current GRASP2018 \cite{grasp2018} version or after minor changes with the upcoming new GRASP version \cite{NewGrasp_2023}.

We are thankful to Chunhai Lyu,  Marianna Safronova,  Sergey Porsev,  and Charles Cheung for useful discussions.  AP gratefully acknowledges funding from the Deutsche Forschungsgemeinschaft (DFG) in the framework of the Heisenberg Program.

\section*{References}
\bibliography{refs_ml_csf}

\end{document}


\title{Deep-learning approach for the atomic configuration interaction problem\\on large basis sets:\\Supplementary Material}

\author{Pavlo Bilous}
\email{pavlo.bilous@mpl.mpg.de}
\affiliation{Max Planck Institute for the Science of Light, Staudtstra{\ss}e 2, 91058 Erlangen, Germany}
\affiliation{Max Planck Institute for Nuclear Physics, Saupfercheckweg 1, 69117  Heidelberg, Germany}
\author{Adriana P\'alffy}
\affiliation{Institute of Theoretical Physics and Astrophysics, University of W\"urzburg, Am Hubland, 97074 W\"urzburg,  Germany}
\affiliation{Max Planck Institute for Nuclear Physics, Saupfercheckweg 1, 69117  Heidelberg, Germany}
\author{Florian Marquardt}
\affiliation{Max Planck Institute for the Science of Light, Staudtstra{\ss}e 2, 91058 Erlangen, Germany}
\affiliation{Department of Physics, Friedrich-Alexander-Universit\"at Erlangen-N\"urnberg, 91058 Erlangen, Germany}

\maketitle

\section{Solution of the CI problem in the CSF basis using GRASP2018}

In the configuration interaction (CI) approach,  the wave function of the considered state $\ket{\Psi}$ is represented as an expansion $\ket{\Psi} = \sum_\alpha c_\alpha \ket{\Phi_\alpha}$ with unknown coefficients $c_\alpha$, where  $\alpha$ stands for all quantities uniquely characterizing the basis state.  The expansion coefficients $c_\alpha$ and the energy $E$ of the electronic state are obtained by solving the diagonalization problem $\hat{H} \ket{\Psi} = E \ket{\Psi}$,  where for atomic systems $\hat{H}$ is usually the Dirac-Hartree-Fock or Dirac-Hartree-Fock-Breit Hamiltonian [2].  In the present work we consider the Dirac-Hartree-Fock Hamiltonian and use the basis of configuration state functions (CSF) [1,2] as the basis set $\{\ket{\Phi_\alpha}\}$.  

For a given electronic configuration $n_1l_1^{a_1} \ldots n_k l_k^{a_k}$ (here $a$ is the occupation of the orbital with the principal and orbital quantum numbers $n$ and $l$, respectively),  CSFs are constructed from individual electronic wave functions by subsequently considering (1) different distributions of electrons over the relativistic suborbitals; (2) all possible angular momenta couplings within each suborbital; (3) all possible couplings between the suborbitals to the total angular momentum of the considered energy level.  As a wave function,  each CSF consists of products of Dirac central field orbitals combined in a way that guarantees the antisymmetry upon electronic permutations and the proper rotational symmetry corresponding to the considered energy level.  This leads to a great advantage over the ordinary basis of Slater determinants consisting in the possibility to separate completely the ``angular'' and the ``radial'' parts of the matrix elements using the Wigner-Eckart theorem [20].  The former part can be addressed using the methods of the quantum angular momentum theory (see Ref.~[21] and references therein),  whereas the latter reduces to evaluation of radial integrals,  i. e.  usual one-dimensional integrals with the radial part of the one-electron wave functions.

A single electronic configuration $n_1l_1^{a_1} \ldots n_k l_k^{a_k}$ yields usually only a few to a few tens of CSFs.  However, for accurate atomic structure data, correlation  effects [2] need to be taken into account by additionally including  CSFs originating from other configurations with the same good quantum numbers of the total angular momentum $J$ and parity $\pi$.  The targeted accuracy and the strength of the electronic correlations determine the size of the required basis set.  These additional configurations are obtained from the initial one by moving (``exciting'') electrons from occupied orbitals to vacancies in partially occupied or empty (virtual) orbitals up to some fixed orbital $n_v l_v$.  In practice, the innermost core electronic shells are kept frozen and only excitations from higher occupied orbitals down to a fixed orbital $n_c l_c$ are allowed.  Due to the very fast growth of basis sets obtained in this way,  often only single (S) and double (D) excitations are considered leading however in some cases to still very large basis sets.  We denote the CSF set constructed in this way as $X(n_c l_c,  n_v l_v)$,  where the generic notation $X$ stands for S,  D or their combination.  For instance, the sets $\mathrm{SD}(3p, 9h)$ for the ground state of $\isotope{Re}$ and $\isotope{Os}$ atoms considered here and in Ref.~[26],  consist of over 90 million CSFs --- a prohibitively large number from the point of view of the needed computational resources.

The General Relativistic Atomic Structure Package GRASP2018 [19] was used in this work for the numerical solution of the CI problem.  In the applied procedure,  the program \texttt{rangular} from the GRASP2018 package performs the angular part of the computation and uses the disk to store the result,  which is then loaded by the program \texttt{rmcdhf} into the memory for multiplication with the radial integrals and further diagonalization.  Note that \texttt{rangular} stores temporarily also  intermediate results leading to a significant growth and then decrease of the used disk space during the program run.  This scenario determines the required disk space and memory,  which are both the bottleneck in large CI calculations with GRASP2018.  For a few million CSFs the typical consumption of the both resources is measured in TB and is still affordable on scientific clusters.  However due to the roughly quadratic growth of the needed space with the CSF set size,  using larger expansions becomes quickly infeasible.  The bases of approx.  90 million CSFs considered in this work and in Ref.~[26] are prohibitively large and can be addressed only in a relatively small part,  which must consist of the most important CSFs.  In this work we have developed an approach which allowed us to achieve the needed energy accuracy on sets of up to 5 million out of the 90 million CSFs.  More information on the computational requirements of GRASP2018 and their improvement in the upcoming new GRASP version can be found in Ref.~[31]. 

\section{ Kernel processing of the input data in the convolutional NN }

The convolutional neural network (CNN) shown in Fig. 1a) in the Main Text has proven to work robustly and yield precise results.  We made sure that no significant improvement could be achieved by varying different parts of the NN.  We show here the study in which the size of the input kernel (B) [see Fig. 1a) in Main Text] was varied.  The kernel processing of the input proceeds via two mechanisms: (a) Application of the same weights along the input neurons (independent of the orbitals ordering); and (b) revealing data mutual dependencies for the neighbouring input neurons (sensitive to the orbitals ordering).  From the results shown in Table~\ref{CNN_ker_Table} we can conclude on the contribution of these mechanisms to the better performance of our CNN in comparison to the usual dense NNs demonstrated in Main Text (see Fig.  3 therein).  Indeed,  in the case when the kernel of size 1 is applied,  only mechanism (a) is present.  From comparison of the final energy precision achieved using a CNN with the input kernel size 1 and the precision obtained using dense NNs shown in Fig.  3  in the main text,  we conclude that mechanism (a) is responsible for most of the advantage yielded by the convolutional architecture.  In this way,  the performance of our method does not show strong dependence on the ordering of the orbitals in the input data.  We use in this work the natural order which is the default option in GRASP2018.  Increasing of the kernel size to $2-3$ leads to a somewhat better performance due to mechanism (b) which however drops when increasing the kernel size further.

\begin{table}[ht!]
\begin{tabular}{|c||c|c|c|c|c||c|}
\hline
Kernel size & 1 & 2& 3 & 4 & 5 & Average  \\\hline
1 & 12.6 & 8.4 &  8.7 &  7.7 &  8.6 & 9.2  \\\hline
2 & 6.6 & 6.7 & 7.3 & 6.6 & 8.4 & 7.1  \\ \hline
\multirow{2}{*}{3}   &   6.4 & 9.5 & 7.1 & 7.2 & 6.4  & \multirow{2}{*}{7.4}     \\\cline{2-6}
						&	8.7  & 6.8 & 7.6 & 6.1  & 7.7 & \\\hline
4 & 8.6 & 8.6 &  6. 1 & 6.3 & 10.9 & 8.1  \\ \hline
5 & 10.6 & 9.1 &  7.9 & 9.6 & 11.0 & 9.6 \\  \hline
\end{tabular}
 \caption{Deviation of the final energy with respect to the ``full'' energy $E ^\text{all}$ in meV obtained in a few computation repetitions using the CNN with different input processing kernel sizes (the data set for kernel size 3 is twice larger than the other ones).  See text for further explanations.}
 \label{CNN_ker_Table}
\end{table}

\section{ Background suppression by the CNN }

\begin{figure}[ht!]
\centering
\includegraphics[width=0.4\textwidth]{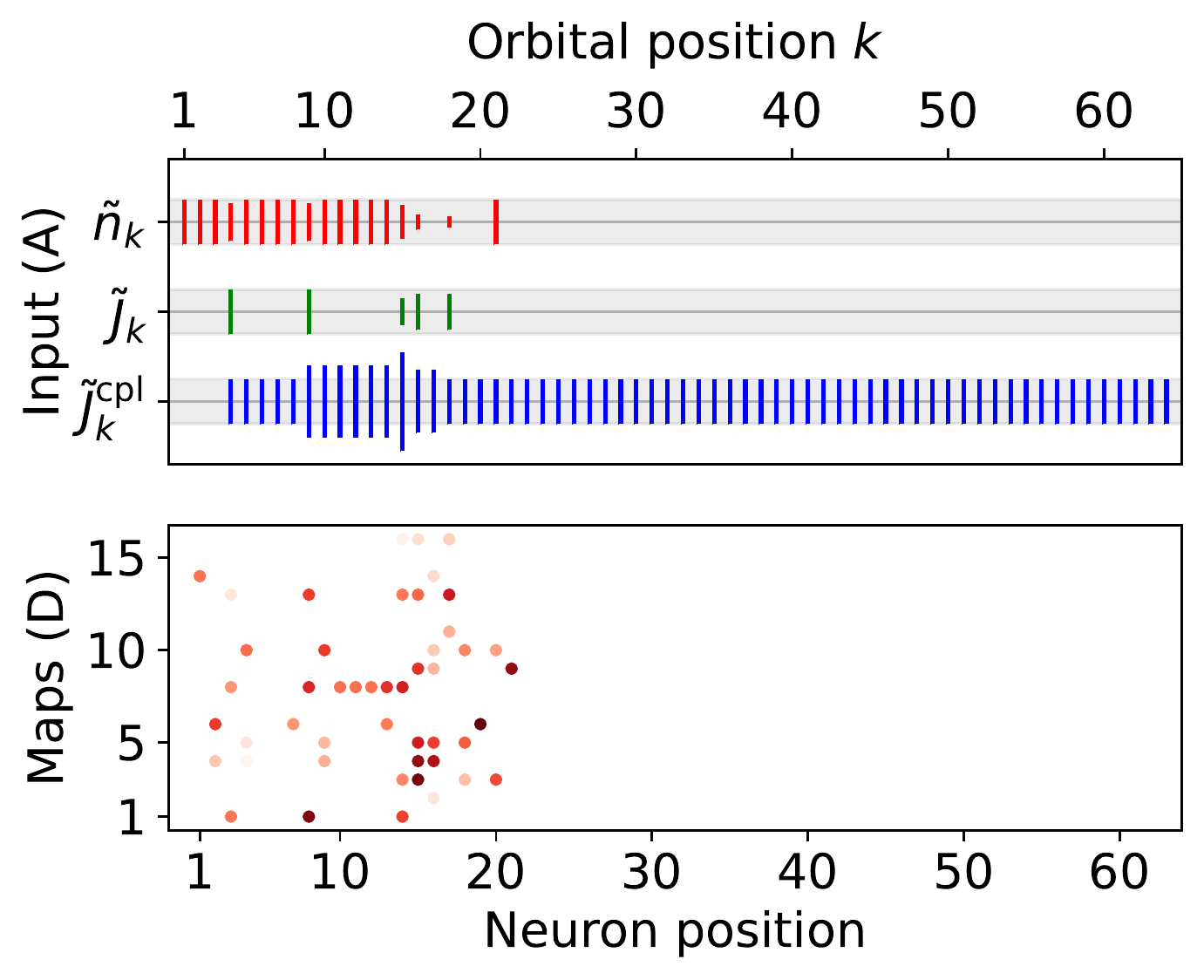}
\includegraphics[width=0.4\textwidth]{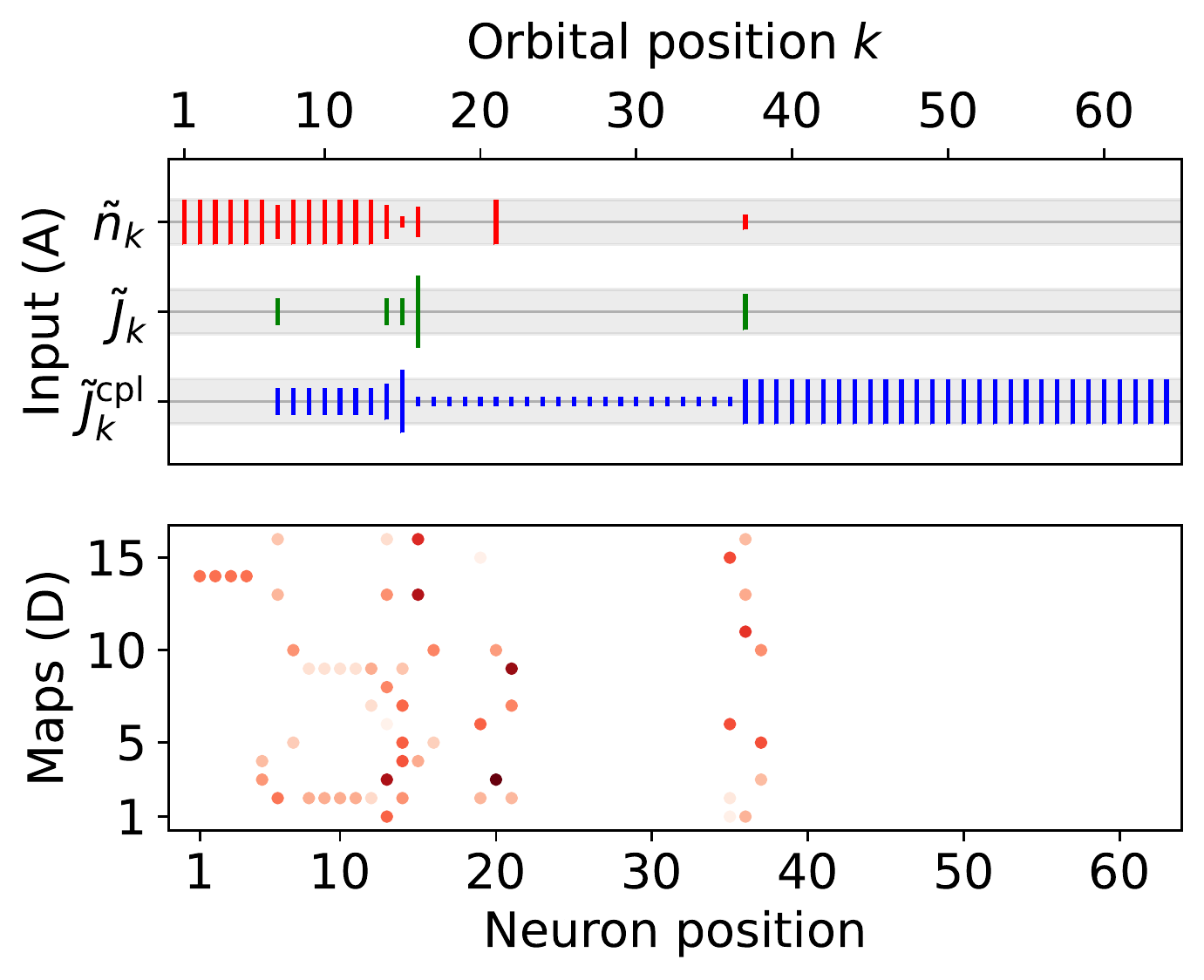}
\includegraphics[width=0.4\textwidth]{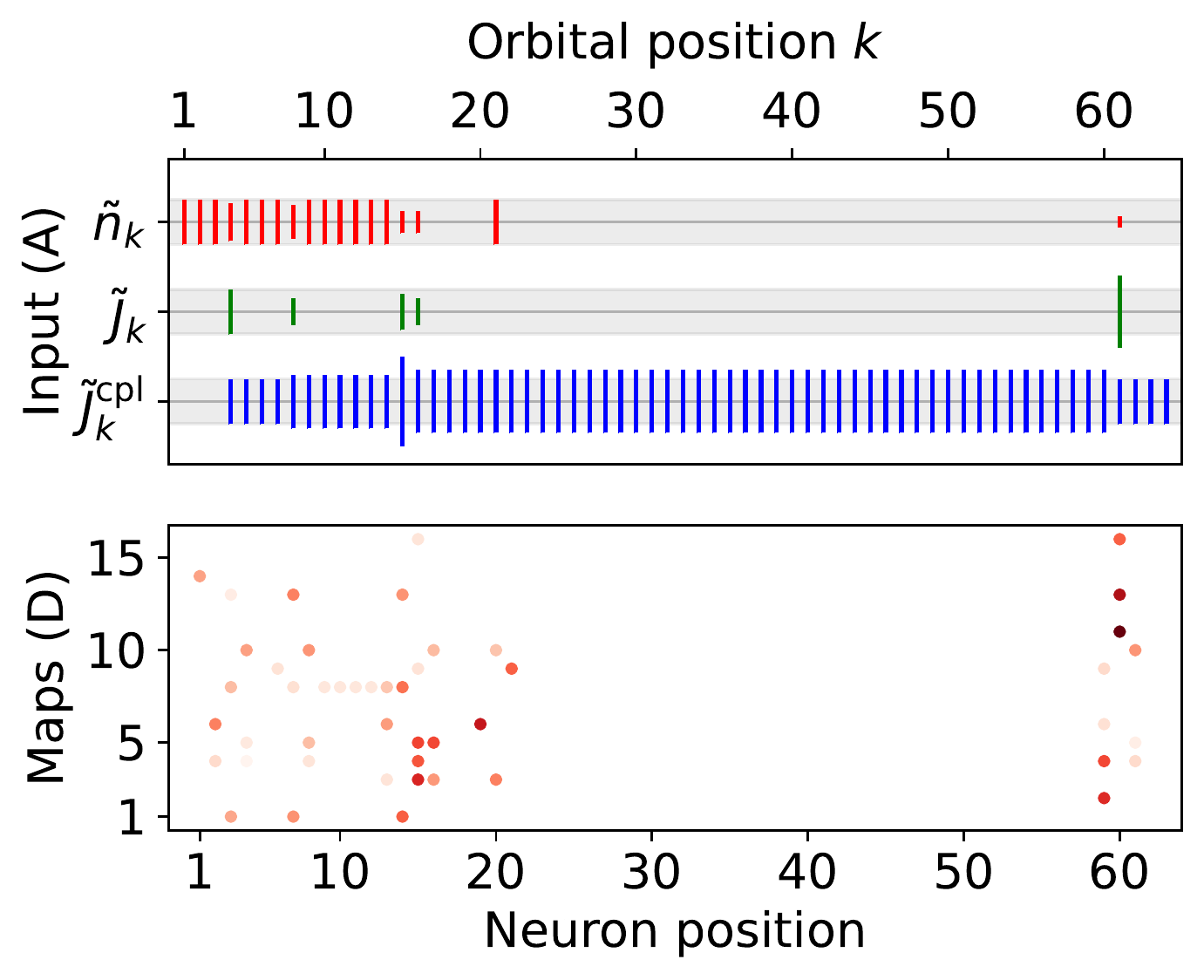}
\includegraphics[width=0.4\textwidth]{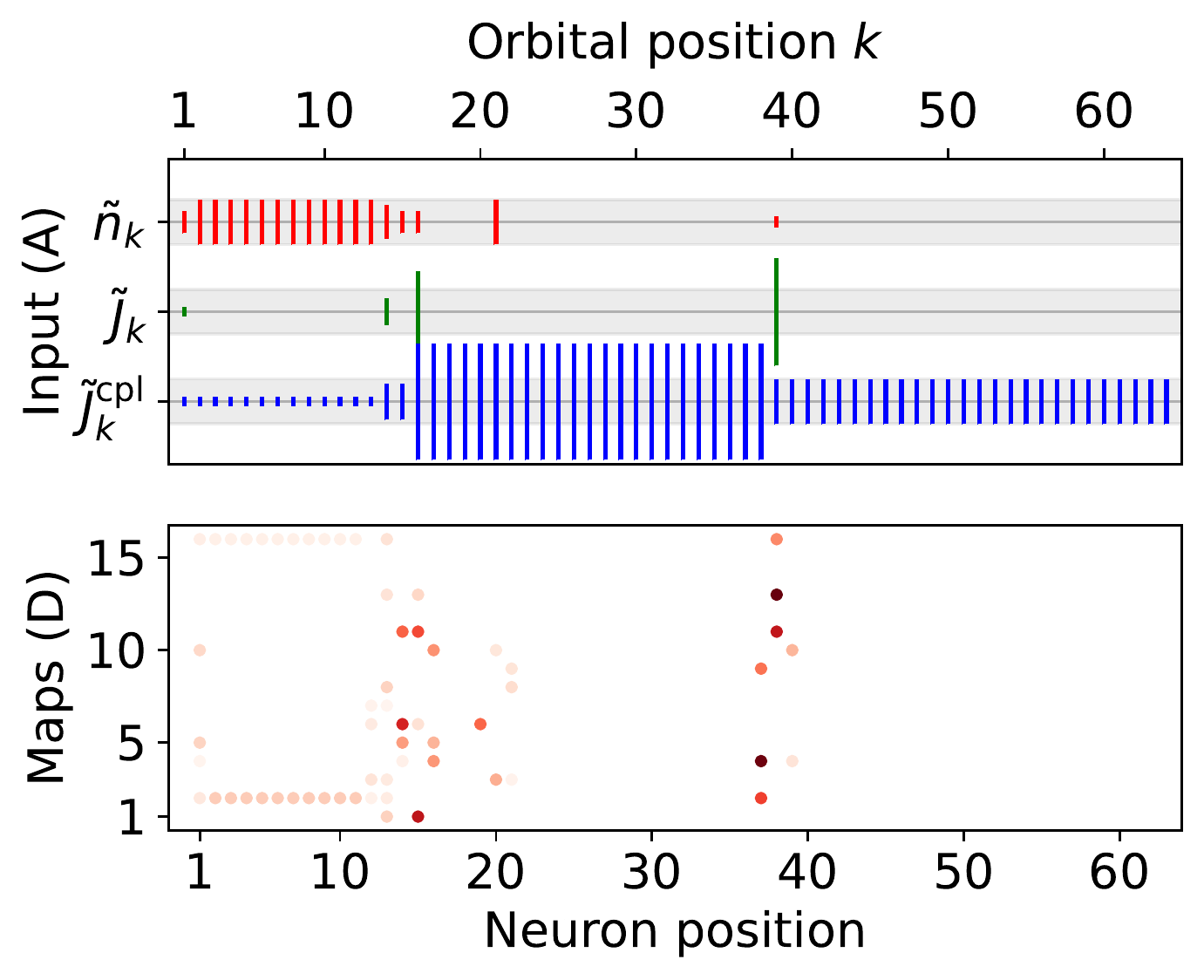}
\includegraphics[width=0.4\textwidth]{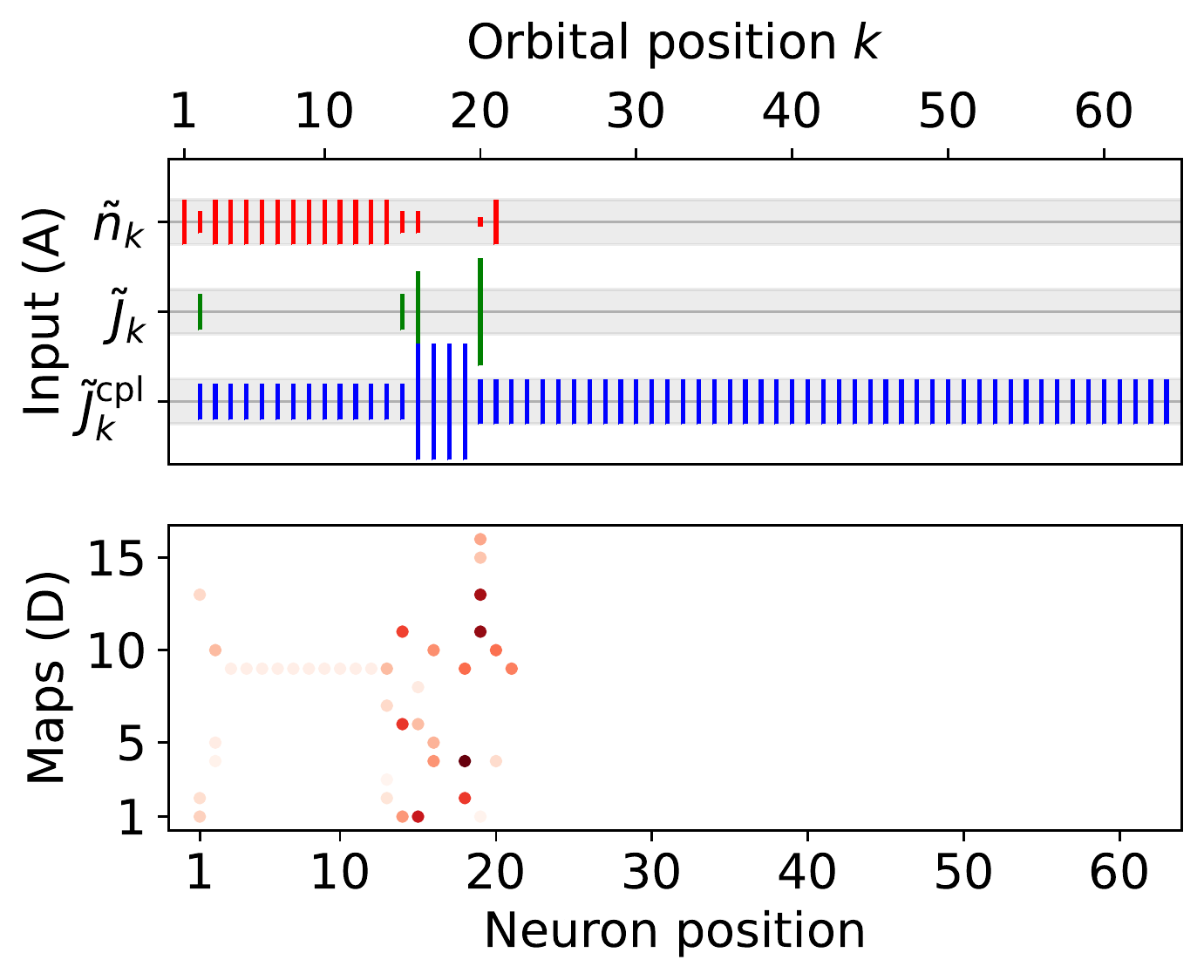}
\includegraphics[width=0.4\textwidth]{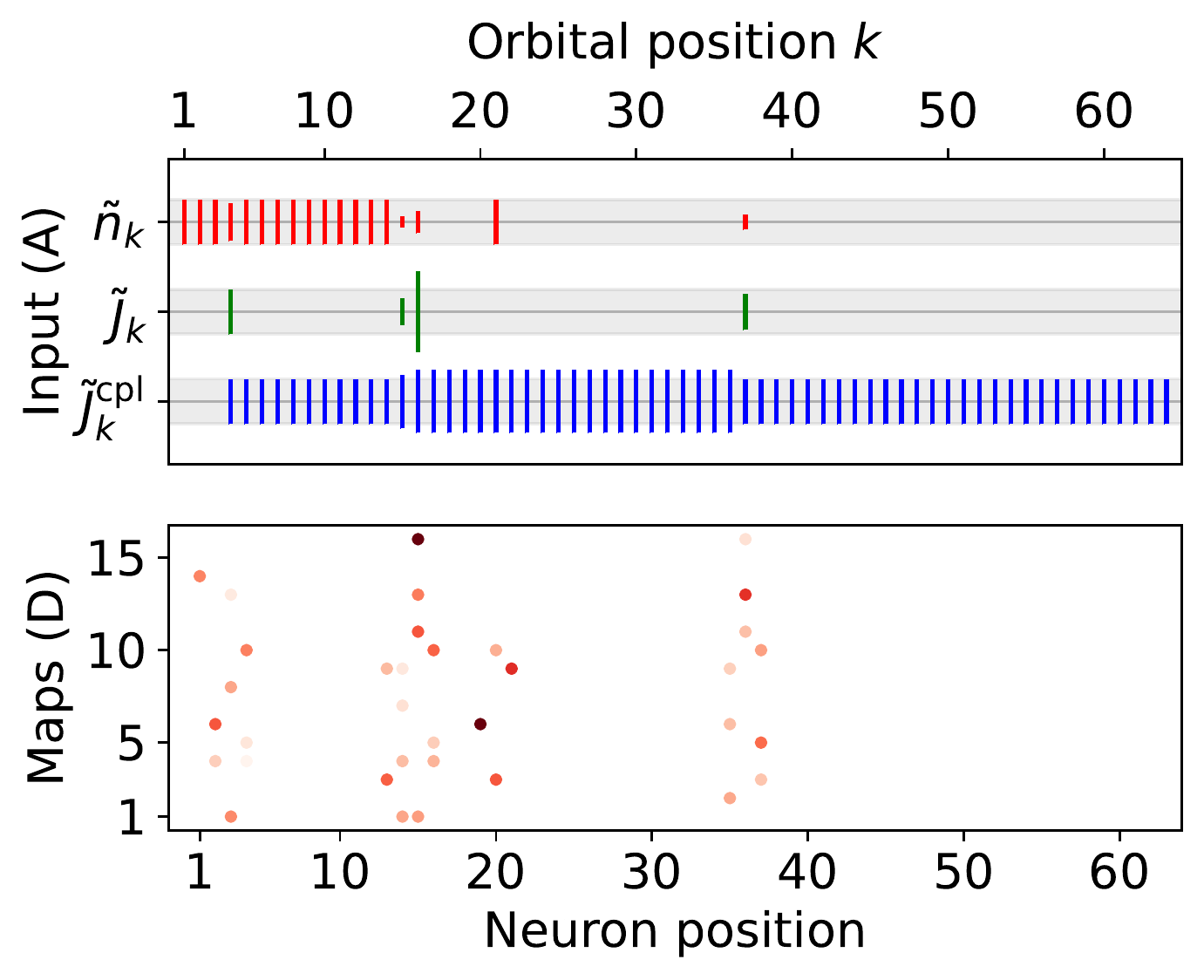}
\includegraphics[width=0.4\textwidth]{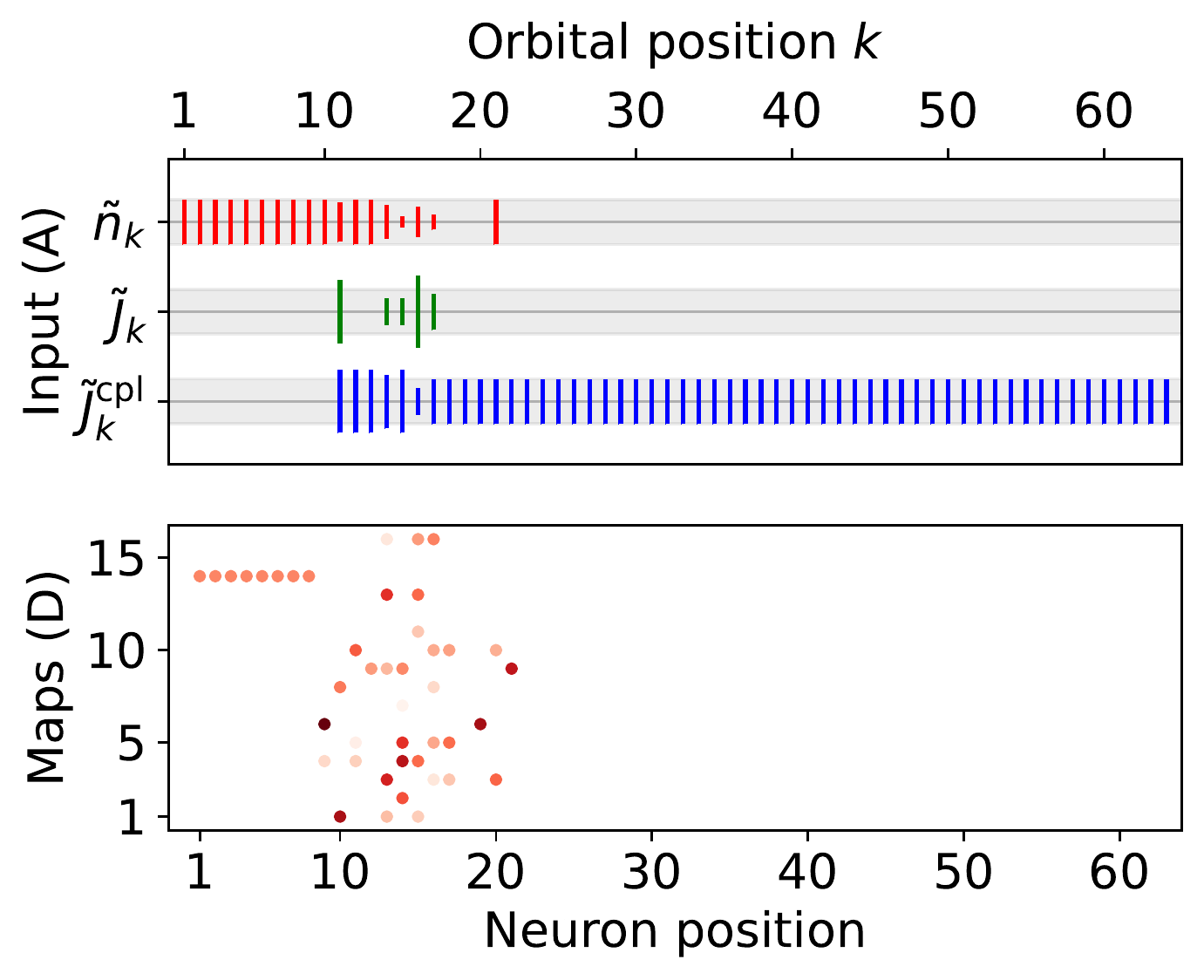}
\includegraphics[width=0.4\textwidth]{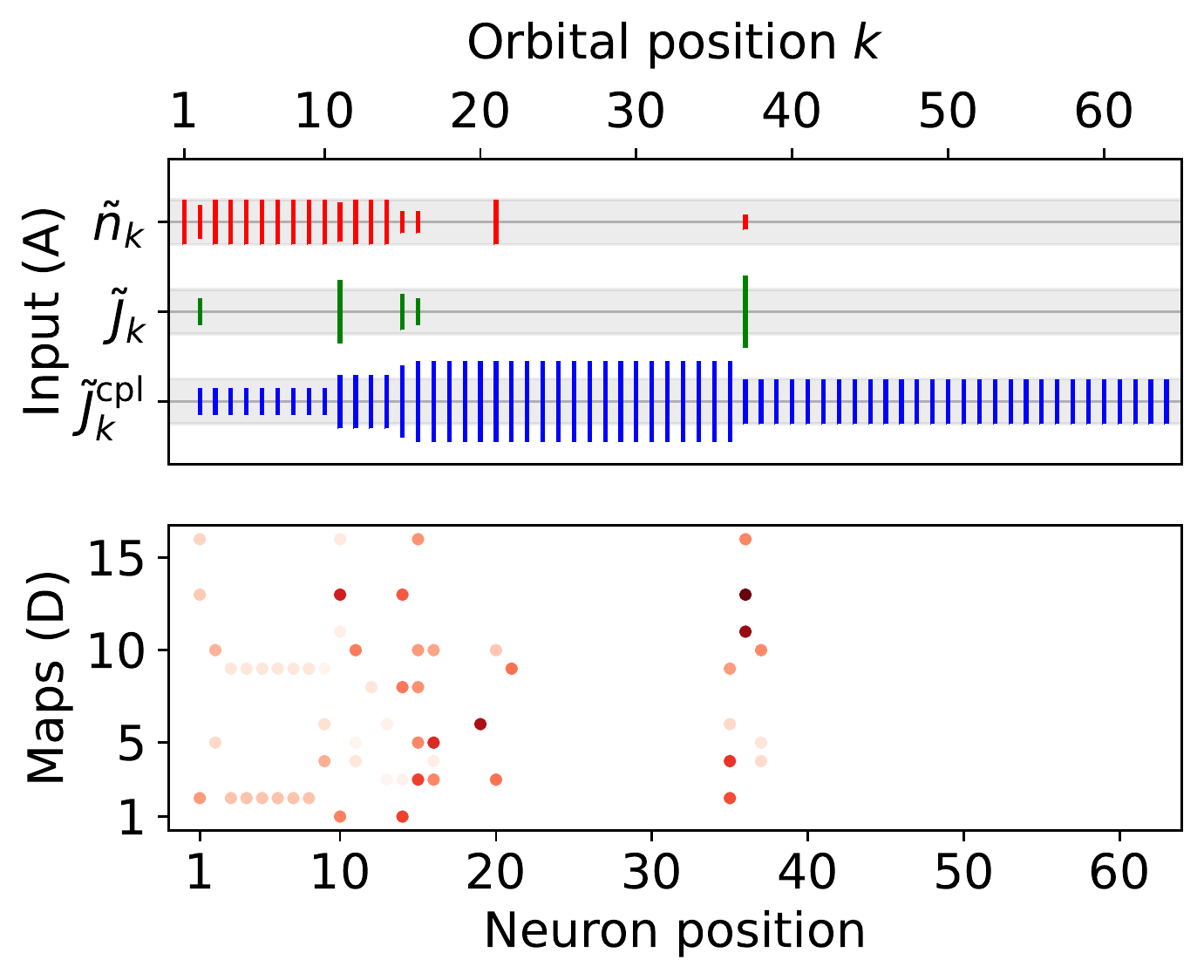}
\caption{(Color online) The color representation and the background suppression effect for 8 randomly chosen CSFs.  See Fig. 1c) and the explanations in the Main Text.}
\label{cnn_imgrec_fig}
\end{figure}

In this Section we provide more information on background suppression by our CNN discussed in Main Text.  In Figure~\ref{cnn_imgrec_fig} we plot the same information as in  Fig. 1c) in Main Text for the same computation but for 8 further randomly chosen CSFs.  As in the Main Text,  the color intensity of the red dots indicates the (always non-negative) values normalized to the maximal value in all the feature maps (higher intensity corresponds to a larger value).  It is seen that the NN tends to always suppress the background corresponding to the completely filled and the vacant orbitals. However,  this suppression is not perfect and residual non-zero values may be present in some neurons in the background region in Fig. 1c).

\begin{figure}[ht!]
\centering
\includegraphics[width=0.4\textwidth]{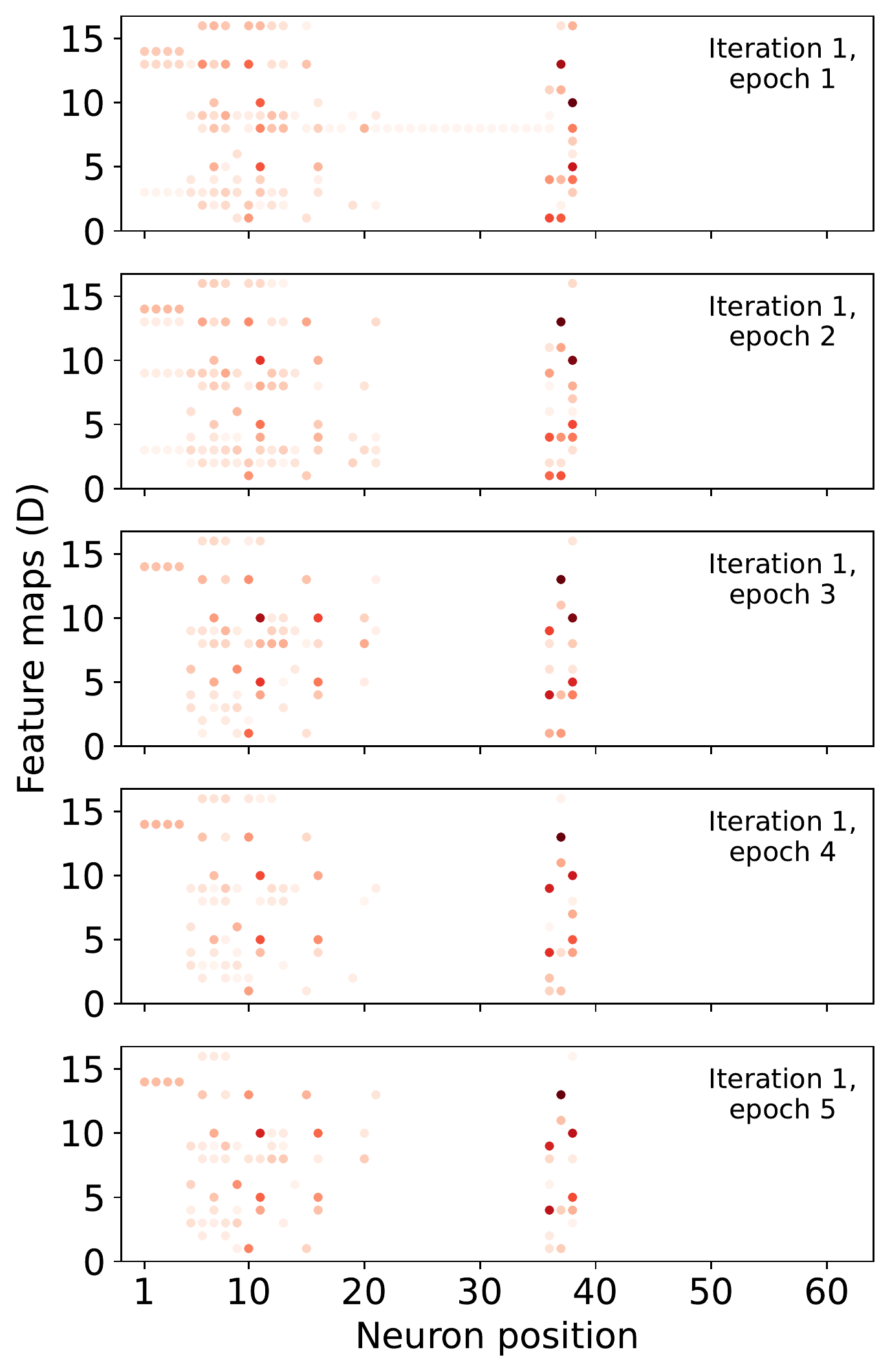}
\includegraphics[width=0.4\textwidth]{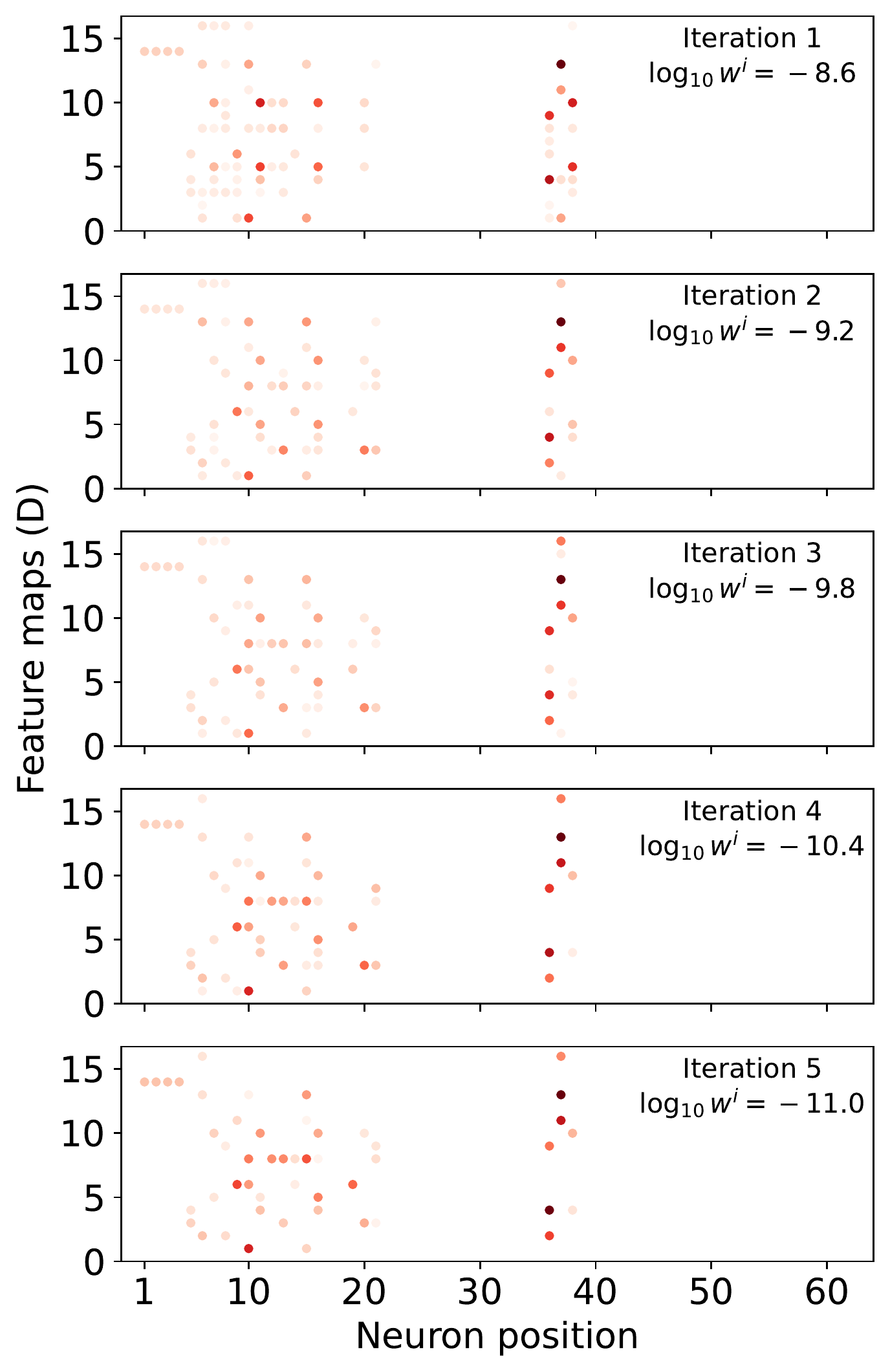}
\caption{(Color online) Neuron values in the feature maps (D) [see the NN architecture in Fig. 1a) in the Main Text] for the exemplary CSF considered in Main Text.  The values are plotted after the first 5 epochs (left panels) of the first iteration and after each iteration labelled by the corresponding cutoff (right panels).}
\label{cnn_imgrec_evol}
\end{figure}

\begin{figure}[ht!]
\centering
\includegraphics[width=0.4\textwidth]{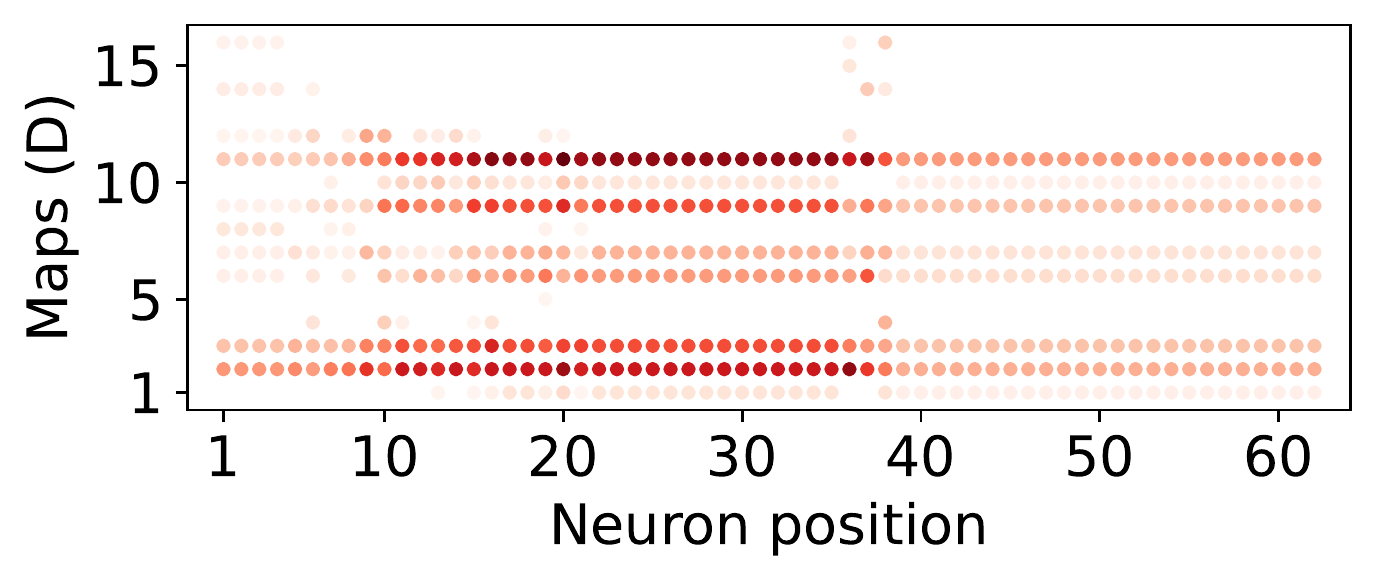}
\includegraphics[width=0.4\textwidth]{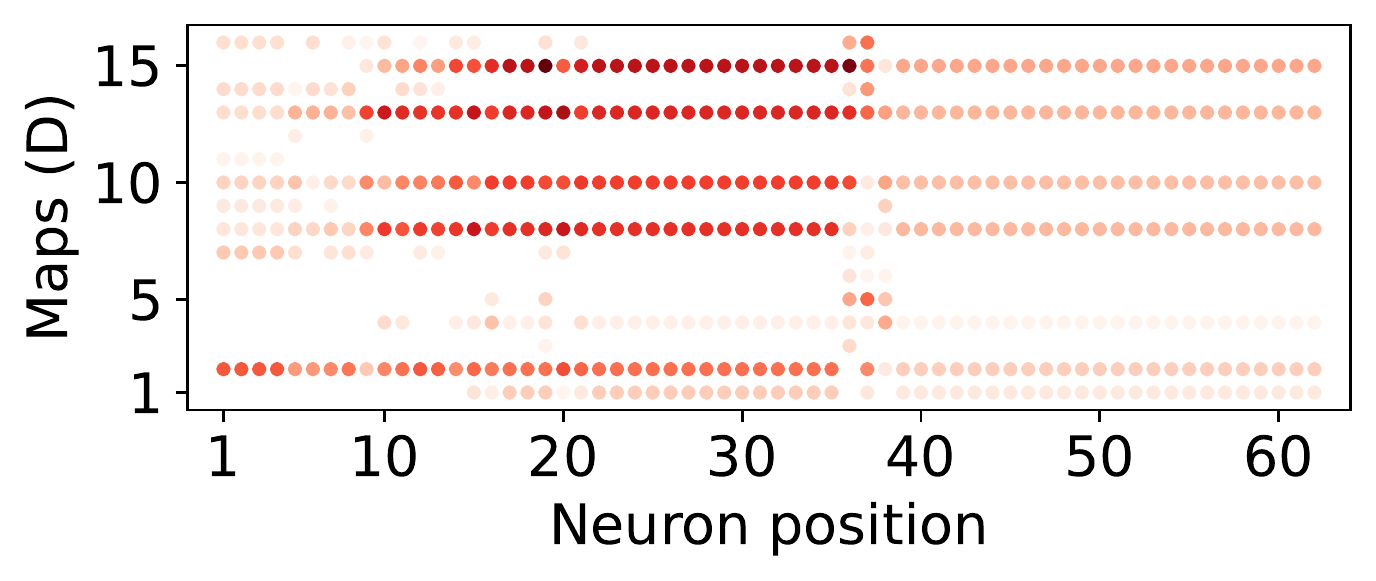}
\includegraphics[width=0.4\textwidth]{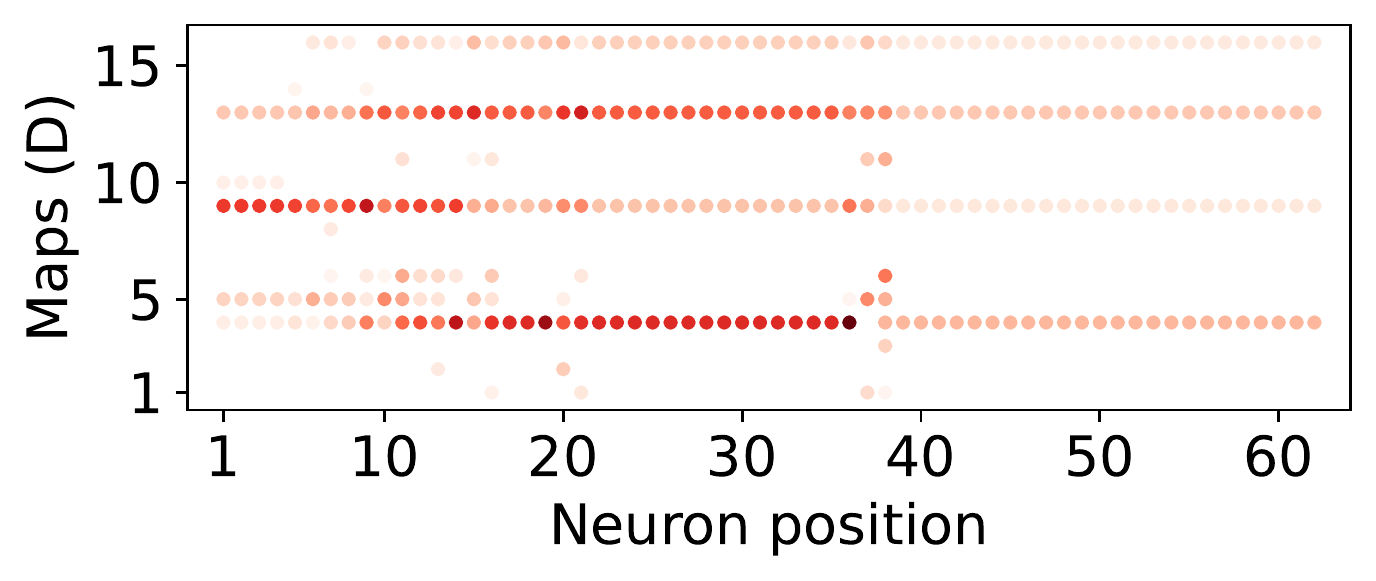}
\includegraphics[width=0.4\textwidth]{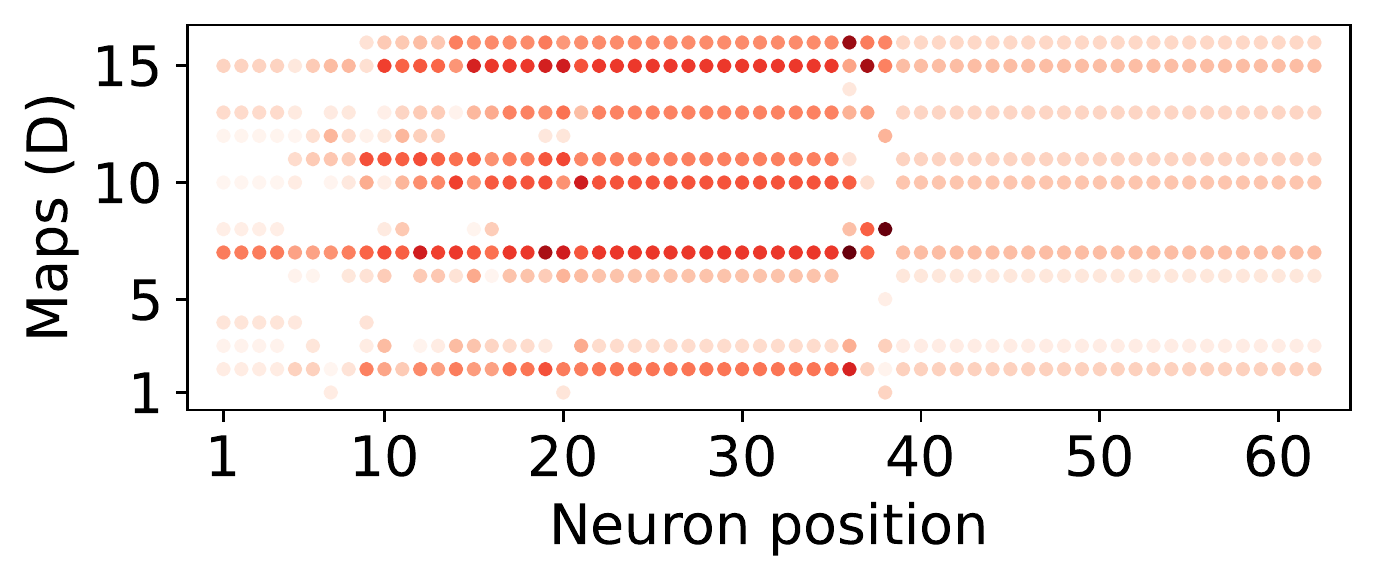}
\includegraphics[width=0.4\textwidth]{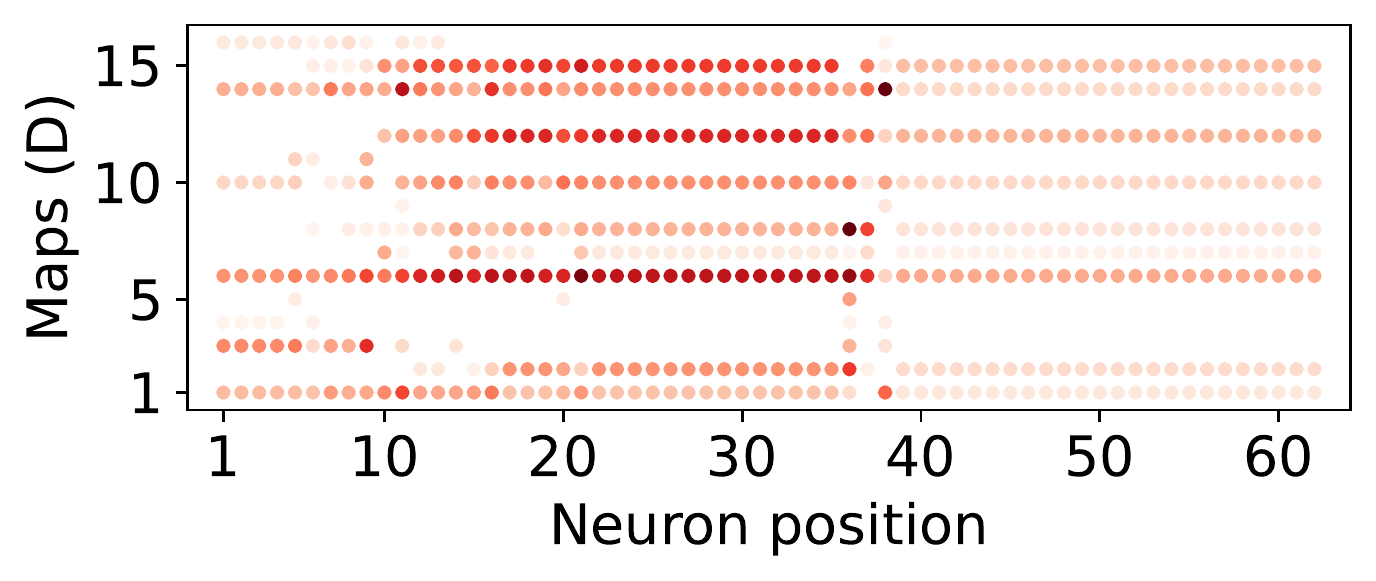}
\includegraphics[width=0.4\textwidth]{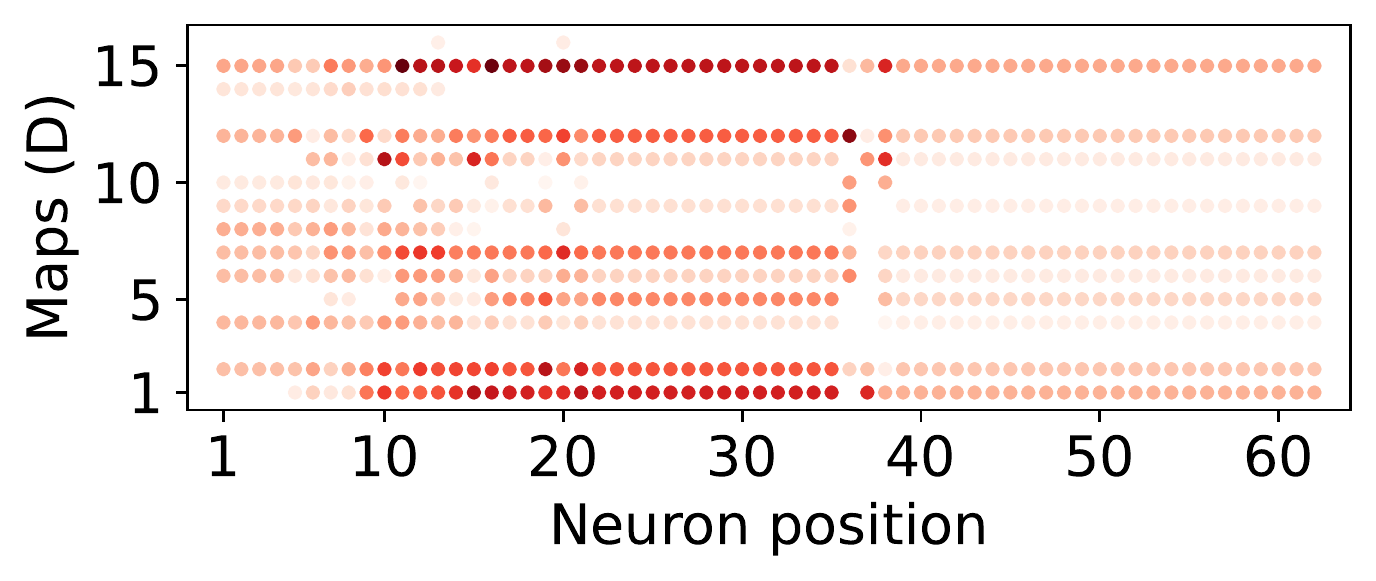}
\caption{(Color online) Neuron values in the feature maps (D) for a freshly created and untrained CNN [see the NN architecture in Fig. 1a) in the Main Text] in 6 independent runs for the exemplary CSF considered in the Main Text. }
\label{cnn_imgrec_init_fig}
\end{figure}

We now switch back to the exemplary CSF considered in Main Text with the color representation shown in Fig.~1b) therein.  During the NN training, the weights in the NN layers are being modified and thus the neuron values on a particular input CSF are changing.   Fig. 1c) Main Text shows the values in the feature maps (D) [see the NN architecture in Fig. 1a) in Main Text] on the considered CSF after the whole training is completed.  Below we show how these values are established during the NN training.  The  computation under consideration contained 5 NN training iterations corresponding to different cutoffs.  These in turn consisted in total of $8 + 16 + 9 + 14 + 13 = 60$ training epochs. Note that since the early stopping approach is used,  the number of epochs in the iterations may vary from run to run.  In the left panels of Fig.~\ref{cnn_imgrec_evol}  we show the state of the feature maps (D) after the first 5 epochs of the first iteration.  We see that the overall background suppression takes place already at the very beginning of the NN training.  The obtained structure is only refined in further epochs,  as seen in the right panels of Fig.~\ref{cnn_imgrec_evol} showing the feature maps (D) after each iteration.  Since the weights of a freshly created CNN are initialized randomly,  the values in the feature maps (D) before training differ significantly from run to run.  In Fig.~\ref{cnn_imgrec_init_fig} we show the state of feature maps (D) before training in 6 runs.  In all these cases the CNN identifies and suppresses the background already in the first training stage.

